\documentclass[twocolumn]{aastex63}
\usepackage{times}
\usepackage{amsmath}
\usepackage{url}
\usepackage{graphicx}
\usepackage{CJK}
\long\def\symbolfootnote[#1]#2{\begingroup%
	\def\thefootnote{\fnsymbol{footnote}}\footnote[#1]{#2}\endgroup}

\newcommand{\beq}{\begin{equation}}
\newcommand{\eeq}{\end{equation}}
\newcommand{\bea}{\begin{eqnarray}}
\newcommand{\eea}{\end{eqnarray}}

\begin{document}
	
\title{\large The first OGLE-discovered ultracompact X-ray binary is an intermediate polar}

\begin{CJK*}{UTF8}{gkai}
	
\author{Shuai Peng (彭帅)}
\affiliation{School of Physics and Astronomy, Sun Yat-Sen University, Zhuhai, 519082, China}
\author{Rong-Feng Shen (申荣锋)}\thanks{shenrf3@mail.sysu.edu.cn}
\affiliation{School of Physics and Astronomy, Sun Yat-Sen University, Zhuhai, 519082, China}

\begin{abstract}
		
The variable source OGLE-UCXB-01 is the first OGLE-discovered ultracompact X-ray binary (UCXB). The 12-year long-term OGLE optical photometry of this source shows a period of $P=12.8$ min and a fast period decreasing rate $\dot{P}= -9.2 \times10^{-11}$ s s$^{-1}$. At a luminosity of $L_X \approx 4 \times10^{33}$ erg s$^{-1}$, its X-ray emission is also variable and correlated with the optical variability. To determine the nature of this variable source, specifically the masses and types of its binary components, we consider first an attractive possibility that the optical variation is due to the secondary's ellipsoidal variation and a strong gravitational wave emission drives the orbital decay. However, we can not find an allowable solution to the secondary that satisfies simultaneously the three constraints: an ultra-tight orbit, the bright absolute magnitude, and the large amplitude of the brightness variation. Moreover, the inferred mass transfer rate is too high. This scenario is therefore ruled out. We then find the system is fully consistent with an ``intermediate polar'' model, in which the optical and X-ray emission comes from a magnetized white dwarf (WD) accreting from a low-mass ($\lesssim 0.7~ M_\odot$) main-sequence secondary. The observed period decay is the accretion-driven spin-up of the WD.  The WD spin period is 12.8 min and the orbital period is shorter than 10 hr. The method presented here can be applied to other UCXB candidates or impostors with time-domain data available only.
		
\end{abstract}
\keywords{Stellar accretion -- White dwarf stars -- Magnetic variable stars -- X-ray binary stars -- Periodic variable stars}

\section{Introduction}
	
	Finding fast ($<$ hr) periodic variable sources is of important values in astrophysics. First, if the brightness variation is due to the orbital modulation in a ultracompact binary, then it would be a potential gravitational wave (GW) source. Second, if the emission is due to accretion of matter, it would be an ideal object to study mass transfer and its role on binary evolution. Lastly, when the variable has an X-ray counterpart, it usually suggests the existence of a compact object, either a white dwarf (WD), neutron star (NS) or a black hole (BH).
	
	OGLE-UCXB-01 is a variable source discovered by \cite{pietru19} in a 12-year long-term Optical Gravitational Lensing Experiment (OGLE) photometry observation. OGLE is a campaign to search for variable stars using a 1.3m telescope \citep{udalski15}. The source is located in the field of Galactic bulge globular cluster Djorg 2 (distance 8.75 kpc). The long-term (2010-2018) photometry in $I$ band shows an average amplitude of variation $\Delta I \approx 0.35$ magnitude. Fourier analysis reveals a period of $P=12.79$ minute and moreover, a constant and fast period decreasing rate $\dot{P}= -9.17(16)\times10^{-11}$ s s$^{-1}$ \citep{pietru19}.
	
	A 6.3-hour archival \textit{Chandra} observation detects a point X-ray source 0$''$.64 away from the position of the variable source, whose spectrum is an absorbed power law with a photon index of $1.22\pm0.23$, with an unabsorbed flux in 0.5-10 keV of $\approx 4.8\times10^{-13}$ erg s$^{-1}$ cm$^{-2}$, which corresponds to a luminosity of $L_X \approx 4.4\pm0.5 \times10^{33}$ erg s$^{-1}$ \citep{pietru19}. The phase-folded X-ray light curve also shows a temporal variation that is clearly correlated with the optical one.
	
	The \textit{HST} photometry of the source measures a $V$ band brightness of 21.2 mag. Adopting an extinction of $A_V \approx 2.4$ in the direction of the cluster Djorg 2, \cite{pietru19} estimated an absolute magnitude of the source $M_V \approx +4.1$.
	
	Identifying this period as the orbital period of a binary, \cite{pietru19} classified it as a ultracompact X-ray binary. Its period shows a steady and fast decrease with time. About a dozen of such sources of this category have been found so far \citep{nelemans10}. Is this period decay mainly due to the gravitational wave (GW) emission from the binary, or by some other mechanism? What are the nature of the binary's components? Answering those questions are the aims of this paper.
	
	OGLE-UCXB-01 appears to be in \textit{Gaia} DR2\footnote{\url{https://gea.csac.esa.int/archive}} (source ID: Gaia DR2 4062733092919727104) and EDR3 (source ID: 4062733092980359040) but with no parallax. This is consistent with its distance being the same as that of Djorg 2, which \cite{pietru19} have assumed and we will follow here.
	
	First, in \S \ref{sec:general} we assume that the source is in a binary, its brightness variation is due to an orbital modulation and the period decay is due to the GW emission. From these, we derive constraints on some observed properties and compare them with the data to see if this scenario is acceptable. Then in \S \ref{sec:polar} we consider an alternative ``intermediate polar'' model, where the source is in a binary with a magnetic WD and an unknown secondary, but the optical variation is \textit{not} modulated by the orbital motion. Our conclusion is given in \S \ref{sec:conc}. 
	
\section{Brightness variation due to binary orbital motion} \label{sec:general}
	
 	In this paper we take a more general approach. First, in this section we follow \cite{pietru19} to assume that the source is in a binary, its brightness variation is due to an orbital modulation and the period decay is due to the GW emission. Then we compare the inference of this scenario with observations. Later we consider an alternative scenario in \S \ref{sec:polar}.
	
	Suppose the total mass of the binary is $M \equiv M_1 + M_2$, where $M_1$ and $M_2$ are the component masses, and the average separation between them is $a$. The orbital period $P_{\rm orb}$ is more or less comparable to the observed brightness variation period $P$. Then from the Kepler's 3rd law $P_{\rm orb}^2= 4\pi^2 a^3/GM$ we have
	\beq		\label{eq:a}
	a =  0.18 \left(\frac{M}{M_{\odot}}\right)^{1/3} \left(\frac{P_{\rm orb}}{12.8 \min}\right)^{2/3} R_{\odot}.
	\eeq 
	
	This tight orbit shown in equation (\ref{eq:a}) makes it very hard for a main-sequence (MS) star or one on its evolved phase to fit in. On the other hand, the source's substantial X-ray emission suggests that it contains at least one compact object, i.e., a white dwarf (WD), neutron star (NS) or black hole (BH). 
	
	\subsection{The secondary fills its Roche-lobe}	\label{sec:roche}
	
	One ordinary way of generating substantial X-ray emission is by the accretion of the mass transferred from the donor to a compact object. This suggests that the donor has just filled its Roche lobe, i.e., the radius of the donor $R_2$ is equal to its Roche lobe radius $R_{L2}$:
	\beq		\label{eq:roche}
	R_2 \simeq R_{L2}=  0.46 \left(\frac{M_2}{M}\right)^{1/3} a.
	\eeq  
	Here we use \cite{paczynski71}'s approximate Roche-lobe size formula for mass ratio $M_2/M_1 \lesssim 1$ (i.e., the donor is the secondary). 
	
\begin{figure}[h]
		\begin{center}
			\includegraphics[width=8cm, angle=0]{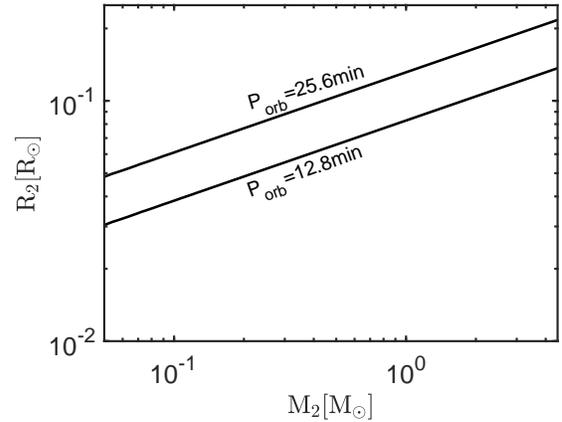}
			\caption{The mass-radius relation of the donor obtained from Kepler's 3rd law Eq. (\ref{eq:a}) and  the Roche-lobe filling condition Eq. (\ref{eq:roche}). Within the scenario that the observed periodic $I$-band brightness variation is modulated by the orbital motion, we consider two values of the orbital period, $P_{\rm orb}= P$ and $2P$, respectively. 
			}    \label{fig:roche}
		\end{center}
	\end{figure}

	Combining Eqs. (\ref{eq:a}) and (\ref{eq:roche}), one gets a relation between the donor's radius and mass, which is plotted in Figure \ref{fig:roche}. This relation is simply the mean mass density of the secondary, i.e., $\rho_2= 110\,({\rm hr} / P_{\rm orb})^2$ g cm$^{-3}$.  
	
	Without going into the detail of the $I$-band brightness modulation mechanism (which we defer to \S \ref{sec:ellipsoidal}), we consider two values for the orbital period: $P_{\rm orb}= P=$ 12.8 mins and $P_{\rm orb}= 2P=$ 25.6 mins, which give $\rho_2 \approx$ 2400 g cm$^{-3}$ and $\rho_2 \approx$ 600 g cm$^{-3}$, respectively.  Typical densities of MS stars are below 100 g cm$^{-3}$. Therefore the secondary is probably a degenerate dwarf. 
	
	If the secondary is a WD, it likely falls into the sub-class of extremely low mass helium WDs ($\lesssim 0.2\, M_{\odot}$) \citep[e.g.,][]{liebert04,kilic07}. Previous mass transfer from it to the primary might explain its low mass.

	\begin{figure}[h]
		\begin{center}
			\includegraphics[width=8cm, angle=0]{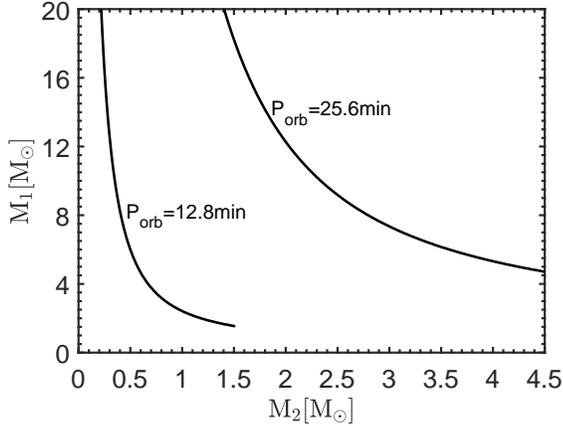}
			\caption{Constraints on the binary component masses from the inferred chirp mass (Eq. \ref{eq:mchirp}) for the scenario that the periodic brightness variation is due to the orbital modulation. The chirp mass is $1.3 ~M_{\odot}$ when $P_{\rm orb}$= $P$= 12.8 mins, and $4.0~ M_{\odot}$ when $P_{\rm orb}$= $2P$= 25.6 mins. The lines in the lower parts stop at $M_1=M_2$.}   \label{fig:mchirp}
		\end{center}
	\end{figure}
		
	\subsection{Gravitational wave emission}	\label{sec:gw}
	
	Here we review the orbital evolution of a binary system, particularly the orbital decay, due to GW radiation \citep{peters64,shapiro04}, since it is an attractive possibility for the case of OGLE UCXB-01. Define the reduced mass of the binary as $\mu= M_1 M_2/M$, and the orbital eccentricity as $e$. The system's orbital energy is $E= -G\mu M / (2a)$. Suppose the orbital decay is due to the GW emission only, then 
	we have
	\beq    \label{eq:Lgw}
	-\left( \frac{dE}{dt} \right)_{GW}= \frac{32}{5} \frac{G^4}{c^5} \frac{M^3 \mu^2}{a^5} f(e),
	\eeq 
	where
	\beq    \label{eq:fe}
	f(e)= \frac{1+ \frac{73}{24}e^2+\frac{37}{96}e^4}{(1-e^2)^{7/2}}.
	\eeq
	So the (time-averaged) orbital period decay rate is 
	\beq    \label{eq:pdot}
	\dot{P}_{\rm orb}= -\frac{192 \pi}{5} \left(\frac{G \mathcal{M}}{c^3}\right)^{5/3}  \left(\frac{2\pi}{P_{\rm orb}}\right)^{5/3} f(e),
	\eeq 
	where $\mathcal{M}= \mu^{3/5} M^{2/5}$ is the chirp mass. In the following we assume the binary orbit is circular, so $f(e)=1$.  
	
	When applying to OGLE UCXB-01, to be more general, we scale $P_{\rm orb}$ with the observed variation period $P$, and express $\dot{P}_{\rm orb}$ accordingly in term of $\dot{P}$. Then plugging the values of $P$ and $\dot{P}$ into equation (\ref{eq:pdot}), we get the chirp mass
	\beq		\label{eq:mchirp}
	\mathcal{M} = 1.3 ~ \left(\frac{P_{\rm orb}}{12.8 \min}\right)^{8/5} M_{\odot}.
	\eeq
	This results in a relation between the binary component masses, which we plot in Figure \ref{fig:mchirp}  for two cases: $P_{\rm orb}= P$ and $P_{\rm orb}= 2 P$ (see below in \S \ref{sec:ellipsoidal} for the motivation of the latter case), respectively.
	It also gives a lower limit (when $M_1 = M_2$) to the binary total mass: $M \geq 3.0~M_{\odot}$.
	
	Given that the secondary mass $M_2 \lesssim 0.2\,M_{\odot}$ from the Roche-lobe filling condition, the primary mass is huge: $M_1 > 20\, M_{\odot}$. According to Eq. \eqref{eq:a}, only a compact object can fit in the tight orbit. Therefore, it seemingly points to a BH as the only solution for the primary.
	
	\subsection{Optical variation due to secondary's ellipsoidal deformation}   		\label{sec:ellipsoidal}
	
	The periodic variation in the $I$-band magnitude may come from the tidal deformation of the secondary, and over one orbit we see different cross sections of the rotating secondary. In this scenario, the orbital period is actually $P_{\rm orb}= 2 P= 25.6$ mins, and its decay rate is twice the observed value as well: $\dot{P}_{\rm orb}= 2 \dot{P}$. From these one can infer that the chirp mass is $\mathcal{M} =4.0 \, M_{\odot}$ (see Eq. \ref{eq:mchirp} and Figure \ref{fig:mchirp}).  
	
	The observed $V$-band absolute magnitude is $M_V \approx 4.1$ \citep{pietru19}, which is more luminous than the Sun ($M_V= 4.83$). Therefore, if we consider that the optical luminosity of the source is dominated by the secondary, the secondary should be a main sequence star. This is already inconsistent with the WD inference from the Roche-lobe overflowing condition in \S \ref{sec:roche}.
		
\begin{figure}
		\begin{center}
			\includegraphics[width=8cm, angle=0]{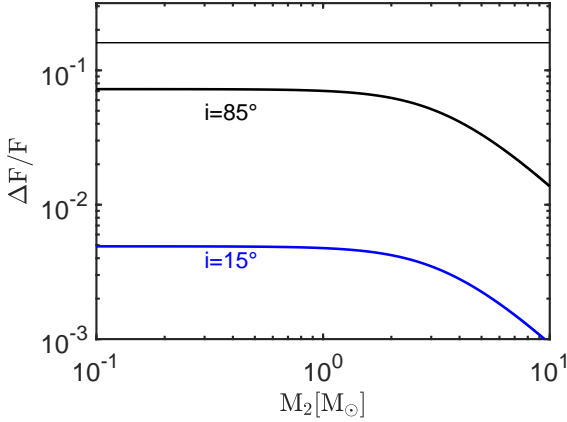}
			\caption{The predicted semi-amplitude of fractional flux variation as a function of the secondary's mass, under the assumption that the periodic flux variation is due to the secondary's ellipsoidal deformation, obtained from the Roche-lobe filling condition (Eq. \ref{eq:roche}), the chirp mass (Eq. \ref{eq:mchirp}) and Eq. \eqref{eq:elli}, where we assumed $u=\tau=0$. The black and blue lines are for two assumed inclination angles, respectively.  The horizontal line represents the observation value $\Delta F/F \approx 0.16$. Note that the secondary's mass is $M_2 < 4.5\,M_{\odot}$ due to the chirp mass constraint (Figure \ref{fig:mchirp}), so the predicted fractional flux variation is too small to match the observation.}    \label{fig:elli}
		\end{center}
	\end{figure}

	The fractional flux variation in this ellipsoidal modulation is estimated by \citep[e.g.,][]{morris93,burdge19}
	\beq \label{eq:elli}
	\frac{\Delta F}{F}= 0.15 \frac{(15+u)(1+\tau)}{3-u} \left(\frac{R_2}{a}\right)^3 \frac{M_1}{M_2} ~\sin^2 i.
	\eeq
	where $u$ is the limb darkening coefficient, $\tau$ is the gravity darkening coefficient and $i$ is the inclination. Plugging the ratio $R_2/a$ from Eq. \eqref{eq:roche} into Eq. \eqref{eq:elli}, we get $\Delta F /F \propto (M_1/M) \sin^2 i$. Then combining with the constraint on the component masses from the chirp mass Eq. \eqref{eq:mchirp}, one reaches a relation between the fractional flux variation and the secondary's mass, which is plotted in Figure \ref{fig:elli}. As it shows, $\Delta F/F$ increases as $M_2$ decreases, until $M_2$ is $\ll M_1$ where $\Delta F/F$ approaches a constant.
	
	The observed semi-amplitude of the $I$-band flux variation is $\Delta F/F\approx 0.16$ \citep{pietru19}, which exceeds significantly the model-predicted values, as is shown in Figure \ref{fig:elli}, unless $M_2$ is small and the system is seen edge-on. Whereas a low-mass ($<$ 0.2 $M_\odot$) WD secondary orbiting a heavy BH ($>$ 20 $M_\odot$) (as implied in \S\ref{sec:gw}) at an edge-on inclination appears to be capable of producing a large-amplitude brightness variation in this scenario, this WD would be too dim to account for the absolute magnitude of the source. 
	
Therefore, the periodic $I$-band variation can not be explained by the tidal deformation of the secondary.

	\begin{figure}
		\begin{center}
			\includegraphics[width=8cm, angle=0]{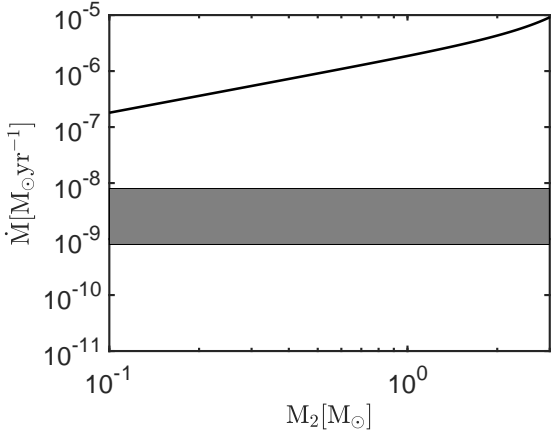}
			\caption{The GW-emission driven mass transfer rate (Eq. \ref{eq:mdot}) as a function of the donor's mass, when using the chirp mass constraint Eq. \eqref{eq:mchirp} and adopting $P_{\rm orb}=25.6$ mins. The horizontal grey strip shows the range of $\dot{M}$ estimated from the observed $L_X$ adopting a radiative efficiency of $10^{-4} \sim 10^{-5}$.}   \label{fig:mdot}
		\end{center}
	\end{figure}

	\subsection{X-ray luminosity due to accretion onto primary}	\label{sec:xray}
	
	If the X-ray emission is due to accretion of the transferred mass toward the primary, then its luminosity of $L_X \approx 4.4 \times 10^{33}$ erg s$^{-1}$, adopting a radiative efficiency\footnote{ The values used here are appropriate for WD accretors, which we would later find (see \S \ref{sec:polar}) to be the most likely type of the primary. More compact accretors such as NSs and BHs would have much higher efficiencies, thus suggest even lower $\dot{M}$ estimates.} of $10^{-5 \sim -4}$, would suggest an accretion rate of $8 \times 10^{-10 \sim -9} M_{\odot}$ yr$^{-1}$. This can be compared with a predicted mass transfer rate under the assumption that the mass transfer is driven mainly by orbital decay due to the GW emission \citep[e.g.,][]{wick94,ramsay00}:
	\beq		\label{eq:mdot}
	\begin{split}
	\dot{M} \approx & 1.7\times 10^{-10} \times \\
	&  \frac{m_1^2 m_2^2}{(m_1-m_2)} \frac{(P_{\rm orb}/5\, {\rm hr})^{-8/3}}{(m_1+m_2)^{1/3}}  \, M_{\odot}\, \mbox{yr}^{-1}.
	\end{split}
	\eeq
	where $m_1= M_1/M_{\odot}$ and $m_2= M_2/M_{\odot}$.
		
	The chirp mass is $\mathcal{M} = 4.0\,M_\odot$ under the ellipsoidal variation assumption (Eq. \ref{eq:mchirp} and \S \ref{sec:ellipsoidal}). Combing it with Eq. \eqref{eq:mdot}, we get the predicted accretion rate as a function of the secondary mass, which is plotted in Figure \ref{fig:mdot}. It shows that the predicted accretion rate $\sim 10^{-6} M_{\odot}$yr$^{-1}$ is too high to be compatible with that inferred from $L_X$, i.e., it would have overproduced the X-ray luminosity by more than one order of magnitude. 

	To summarize, we consider the scenario of secondary's ellipsoidal variation very unlikely, because it can not simultaneously satisfy these constraints implied by observational facts: an ultra-tight orbit, the bright absolute magnitude, and the large amplitude of the brightness variation (\S\ref{sec:roche} - \S\ref{sec:ellipsoidal}). Moreover, the GW emission is unlikely the cause of the observed period decay, because it would have severely overproduced the accretion rate, i.e., the X-ray luminosity (\S \ref{sec:xray}).  
		
\section{``Intermediate Polar'' model}		\label{sec:polar}
		
	Now we consider an alternative, ``intermediate polar'' model for OGLE UCXB-01, which was proposed for other period-decaying UCXBs \citep[e.g.,][]{ramsay00,roelofs10,strohmayer02}.  In this model, the mass stream from a Roche-lobe overflowing secondary reaches a magnetic WD as the primary, and lands on the WD's magnetic pole, where the X-ray and optical emission is produced and modulated by the rotation of WD. The observed flux variation period $P= 12.8$ mins is actually the spin period of the magnetic WD, which is being spun up by the accretion. The real orbital period of the binary might be much larger, $P_{\rm orb} \sim$ hours, such that even a main-sequence star can be the secondary. 
		
	\subsection{Primary's spin-up due to accretion}
	
	Within the intermediate polar (IP) model \citep[see][for a review]{patterson94}, the transferred mass from the donor forms an accretion disk around the magnitized WD primary. The inner region of the disk truncates at the WD magnetospheric radius \citep[e.g.,][]{ghosh79}
	\beq
	R_m \simeq 0.5\, \mu^{4/7} (2GM_1)^{-1/7} \dot{M}^{-2/7},
	\eeq
	inside of which the disk material is channelled along the curved magnetic field lines onto the primary's surface. Here $\mu \sim B r^3$ is the dipole magnetic moment of the primary, with values of $10^{32 - 34}$ G cm$^3$ for known IPs \citep{patterson94}. 
	
	The primary's angular momentum $J= 4 \pi M_1 R_1^2/(5 P)$ is being increased due to accretion at a rate 
	\beq
	\dot{J}= \frac{4}{5} \pi M_1 R_1^2 \frac{\dot{P}}{P^2} =  \dot{M} \sqrt{GM_1 R_m},
	\eeq
	where for the second equality we assume that the specific angular momentum of the accreted material takes the Keplerian value at $R_m$ \citep{ghosh79}. From this and plugging in the observed $P$ and $\dot{P}$, one can infer the accretion rate as
	\beq    \label{eq:spinup}
	\begin{split}
	\dot{M}= & 5.0\times10^{-9}\,  \mu_{33}^{-1/3} \\ 
	&  \times \left(\frac{M_1}{0.6 M_{\odot}}\right)^{2/3} \left(\frac{R_1}{0.01 R_{\odot}}\right)^{7/3} M_{\odot} \, \text{yr}^{-1}, 
	\end{split}
	\eeq
    where $\mu_{33}= \mu/ 10^{33}$ G cm$^3$.	
	
\begin{figure}[h]
		\begin{center}
			\includegraphics[width=8cm, angle=0]{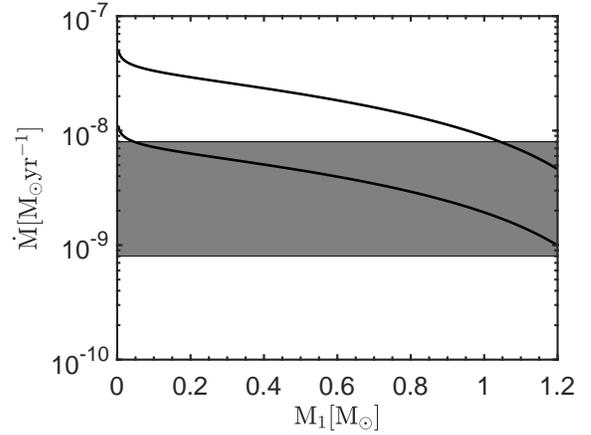}
			\caption{ The primar's mass accretion rate, derived from the primary's spin-up (Eq. \ref{eq:spinup}) within the IP model and the mass-radius relation (Eq.  \ref{eq:mass_radius}) for the WD primary, as a function of the primary mass. The upper and lower curves are for $\mu_{33}=$ 0.1 and 10, respectively. The grey strip area represents the accretion rate estimated from $L_X$ (see \S \ref{sec:xray}), same as in Figure \ref{fig:mdot}.}   \label{fig:mtransfer_primary_conclusion}
		\end{center}
	\end{figure}

	For the primary, we use the mass-radius relation of WDs \citep{Nauenberg1972}
	\beq    \label{eq:mass_radius}
	\frac{R_{\rm wd}}{R_{\odot}}=  \frac{0.0225}{\mu_e} \left( \frac{M_{\rm Ch}}{M_{\rm wd}}\right)^{1/3}
	\left[1-\left(\frac{M_{\rm wd}}{M_{\rm Ch}}\right)^{4/3} \right]^{1/2},
	\eeq
	where $M_{\rm Ch}=5.816/\mu^2 M_{\odot}$ and $\mu_e=2$. Combining Eqs. \eqref{eq:spinup} and \eqref{eq:mass_radius}, we can infer the accretion rate $\dot{M}$ as a function of the primary mass $M_1$, which is plotted in Figure \ref{fig:mtransfer_primary_conclusion}.
	
	Assuming the dipole magnetic moment values $\mu_{33}= 0.1 \sim 10$, the inferred $\dot{M}$ partially overlap with the accretion rates estimated from $L_X$, as is shown in Figure \ref{fig:mtransfer_primary_conclusion}. This suggests a consistency with the IP model. Also note that values of accretion rate $\dot{M} \sim 10^{-9 \sim -8} \, M_\odot$ yr$^{-1}$ inferred from $L_X$  are similar to those of known IPs  \citep[see][]{patterson94}. 
	
	 This consistency gains more weight when we consider the radiative efficiency's dependence on the primary's mass. The X-ray luminosity due to accretion toward the WD primary can be crudely estimated as $L_X \approx GM_1\dot{M}/R_1$ \citep[also see][]{patterson94}. Considering the primary's mass-radius relation (Eq. \ref{eq:mass_radius}), for a given $L_X$, the accretion rate would decrease with $M_1$ (due to an increasing efficiency). This would match qualitatively with the trend of accretion rate inferred by Eq. \eqref{eq:spinup}, shown in Figure \ref{fig:mtransfer_primary_conclusion}.
	 		
	\begin{figure}
		\begin{center}
			\includegraphics[width=8cm, angle=0]{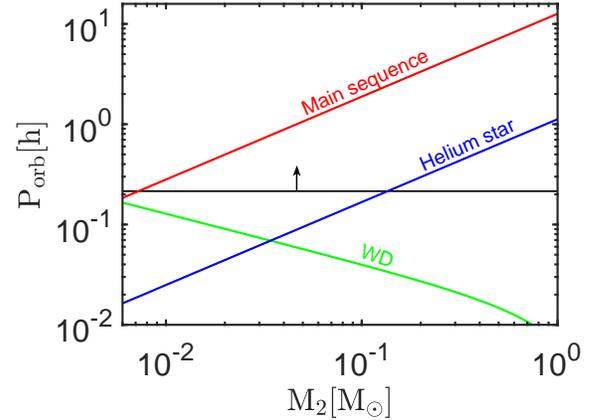}
			\caption{The relation between the secondary's mass and the orbital period, obtained from the orbital separation Eq. \eqref{eq:a}, the Roche-lobe filling condition Eq. \eqref{eq:roche} and the mass-radius relations of WDs, MS and helium stars, respectively, as the secondary. The horizontal line with an arrow represents the primary's spin period $P=12.8$ mins within the IP model, so the orbital period must be larger than this value. Therefore, WDs can be ruled out for the secondary, but the most of MS stars and some higher-mass helium stars are allowed.   }    \label{fig:period_mass}
		\end{center}
	\end{figure}

	\begin{figure}
		\begin{center}
		\includegraphics[width=8cm, angle=0]{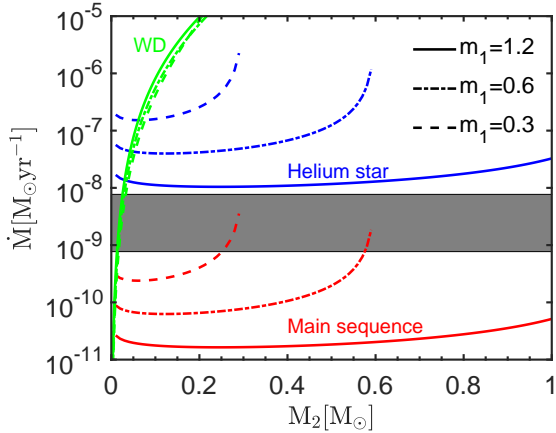}
		\caption{ The mass transfer (accretion) rate as a function of the secondary's mass, for three typical masses of the WD primary, obtained by assuming it is driven by the GW emission (Eq. \ref{eq:mdot}). Other constraints used are the orbital separation Eq. \eqref{eq:a}, the Roche-lobe filling condition Eq. \eqref{eq:roche}, and the mass-radius relations of WDs, MS and helium stars as the secondary, respectively. The grey horizontal strip area represents the accretion rate estimated from $L_X$. It shows that low-mass MSs and low-mass WDs are allowed types of secondary, while helium stars are disfavorred.}    \label{fig:accretion_mass_conclusion}
   		\end{center}
    \end{figure}
	
	\subsection{ Types of the secondary}
	
	In the rest of this section, we determine what types of object are possible for the secondary, within the IP model. We consider three types: WDs, MSs, and helium stars. 
	\begin{itemize}
	
	\item For WDs, we use again Eq. \eqref{eq:mass_radius} as their mass-radius relation. 
	
	\item For MSs, their mass-radius relation can be written as $R_2/R_\odot = (M_2/M_\odot)^{0.88} $ for $ 0.08<$ $M_2/M_\odot$ $<1$ \citep{pont2005}. 
	
	\item Helium star is an evolved stage of a post-MS star whose hydrogen envelope was completely ejected or stripped, for which we use the mass-radius relation: $R_2/R_{\odot}= 0.20\, (M_2/M_{\odot})^{0.88}$ (\citealt{iben91}; \citealp[also see][]{brooks15}).
	
	\end{itemize}
		
	Combining the secondary's mass-radius relations, the orbital separation Eq. \eqref{eq:a} and the Roche-lobe filling condition Eq. \eqref{eq:roche}, we get relations between the secondary's mass and the orbital period, which are plotted in Figure \ref{fig:period_mass}. Although the value of $P_{\rm orb}$ is unknown, we expect it to be larger than the primary's spin period $P= 12.8$ mins. Thus, as are shown in Figure \ref{fig:period_mass}, WDs can be ruled out for the secondary, but the most of MS stars and some higher-mass helium stars are allowed.

	Next, we turn to the accretion rate again. Within the IP model, the GW-driven mass transfer rate Eq. \eqref{eq:mdot} still applies. Combining the orbital separation Eq. \eqref{eq:a}, the Roche-lobe filling Eq. \eqref{eq:roche} and the above mass-radius relations for the secondary, we can express the unknown $P_{\rm orb}$ in terms of $M_1$ and $M_2$. Then plugging it into Eq. \eqref{eq:mdot}, we can plot the mass transfer rates as a function of $M_2$, for three typical values of the WD primary's mass and for the three types of secondary, respectively, shown in Figure \ref{fig:accretion_mass_conclusion}. 	
	
	Then we compare this mass rate with the accretion rate estimated from $L_X$ (\S \ref{sec:xray}), shown as the horizontal strip in Figure \ref{fig:accretion_mass_conclusion}.  It shows that the secondary can not be helium stars, or WDs with mass above $\sim 0.05\, M_\odot$, because they are too compact such that the Roche-lobe filling condition would require a very tight orbit, which causes strong GW emission and drives the mass transfer rate too high. Extremely low mass WDs seem to be allowed but they are already ruled out in Figure \ref{fig:period_mass} because they are still too dense to have a $P_{\rm orb}$ longer than 12.8 mins. 

	Only the low-mass ($< \sim 0.7\, M_\odot$) MS stars seem to satisfy constraints in both Figures \ref{fig:period_mass} and \ref{fig:accretion_mass_conclusion} to be the secondary, on a loose condition that the WD primary is not too heavy ($\lesssim 1\,M_\odot$). In addition, a lower limit to the MS secondary mass $M_2 > \sim 0.01\, M_\odot$ is suggested in Figure \ref{fig:period_mass}.
	
	One last check is that the secondary's absolute $V$ magnitude must be dimmer than the observed mean value $M_V \approx 4.1$, because within the IP model the observed optical brightness variation comes from the accretion onto the spinning magnetized primary. We use the luminosity-mass relation $L_2/L_{\odot}=\left(M_2/M_{\odot} \right)^{4.84} $ \label{eq:mass-luminosity relation} for low-mass ( 0.4 $<M_2/M_{\odot} <$ 1.1) MS stars \citep{eker15} to find a constraint $M_2 < 1.2\, M_\odot$, which is easily satisfied.

\section{Conclusion and Discussion}		\label{sec:conc}
	
	OGLE-UCXB-01 is a periodic optical and X-ray variable with a short period of 12.8 mins. Its constant and fast period decaying rate raises interest in an attractive possibility, in which the strong gravitational wave emission of a compact binary drives the period decay. 
	
	To investigate the nature and the physical parameters of OGLE-UCXB-01, we consider first a scenario that the optical emission comes from the secondary which is tidally deformed by the compact-object primary and undergoes ellipsoidal variation. However, by deriving physical constraints we rule out this scenario, because: 1) To fit into the tight orbit, the secondary which just overflows its Roche lobe must be a WD, but such a dwarf object can not match the observed bright absolute magnitude; 2) The ellipsoidal variation is found to be too small to match the observed amplitude of optical variation for most of the cases; 3) The inferred mass transfer rate and its accretion toward the primary would overproduce the X-ray emission.   
	
	We then consider the IP model, in which a spinning, magnetized WD accretes mass from a secondary; the accretion disk truncates at the WD magnetosphere, inside of which the mass flows along the curved magnetic lines toward the magnetic pole in an azimuthally extended, arced ``curtain'' shape. The heating at or above the WD surface at the pole produces the X-ray / UV / optical emission, which is modulated at the WD spin period \citep[e.g.,][]{patterson94}. The periodic variations in optical and X-rays of OGLE-UCXB-01 are in-phase \citep{pietru19}. This is consistent with the model.\footnote{\cite{Norton04} argue that when the mass from the secondary feeds the magnetic WD in a stream without forming an accretion disk, the optical and X-ray variations could be anti-phased, which might be the case for RX J0806.3+1527 and RX J1914.4+2456.}
	
	Within the IP model, we consider three types of the secondary: low-mass WD, MS star and helium star, respectively. Combing the requirements that the orbital period be larger than the primary's spin period and the mass transfer rate be not much different from the accretion rate estimated from $L_X$, we ruled out WDs and helium stars as the secondary, while a low-mass ($\lesssim 0.7\,M_\odot$) MS star remains as a viable solution for the secondary. 
	
	Under this solution, a constraint on the unknown orbital period, $P_{\rm orb} \lesssim$ 10 h, can be inferred from Figure \ref{fig:period_mass}. This allows us to get the average separation $a \lesssim 2.5 R_{\odot}$ from Eq. \eqref{eq:a}. With these numbers for the orbital parameters, we find that this system is very similar to those confirmed IPs (see Fig. 11 of \cite{patterson94} and this webpage\footnote{\url{https://asd.gsfc.nasa.gov/Koji.Mukai/iphome/iphome.html}}).
	
	A well known IP is AR Scorpii, which is a WD - M dwarf binary with an orbital period of $3.6$ hr and an extremely short spin period of $2$ min \citep{marsh16}. Unlike OGLE-UCXB-01, its X-ray luminosity is extremely low, $L_X \simeq 5\times10^{28}$ erg s$^{-1}$, only 4\% of its optical luminosity, and is $< 1\%$ of the X-ray luminosity of a typical IP. This suggests an extremely weak or no accretion at all toward the WD. In addition, the magnetic WD of AR Scorpii is spinning down on a $10^7$-yr time scale \citep{marsh16}. All these suggest that the multi-wavelength emission of AR Scorpii is spin-powered, whereas OGLE-UCXB-01 is probably accretion-powered.
	
	Future observations can help further confirm the IP origin of OGLE-UCXB-01. For instances, radial velocity measurements can reveal its orbital period, and a high-cadence optical polarimetry could possibly verify that the 12.8-min period is indeed the spin period of the magnetic WD.

	Here we provide a quick estimate of the GW emission from OGLE-UCXB-01. The dimensionless GW strain for a circular binary is \citep[e.g.,][]{Yu10} $h=5\times 10^{-22} (\mathcal{M}/M_{\odot})^{5/3} (P_{\rm orb}/\mbox{hr})^{-2/3} (d/\mbox{kpc})^{-1}$. Taking the mass of primary WD $M_1=0.6\,M_{\odot}$ and the secondary MS $M_2=0.4\, M_{\odot}$, the chirp mass would be $\mathcal{M} =0.42\,M_{\odot}$. We then find the orbital period $P_{\rm orb} \approx 5.9$ h from Figure \ref{fig:period_mass}. So we get the GW emission frequency $f=2/P_{orb}=9 \times 10^{-5}$Hz and $h \approx 4 \times 10^{-24}$. It barely lies at the lower end of the working frequency range $0.1-100$ mHz of the Laser Interferometer Space Antenna\footnote{\url{http://www.srl.caltech.edu/~shane/sensitivity/}} and the TianQin mission \citep{Luo16}, but the signal is too weak for detection.
	
	\acknowledgments
The authors would like to thank the anonymous referee for the valuable feedback and constructive comments. This work is supported by the National Natural Science Foundation of China (12073091), Guangdong Basic and Applied Basic Research Foundation (2019A1515011119) and Guangdong Major Project of Basic and Applied Basic Research (2019B030302001).
	
	
\end{CJK*}
\end{document}